\documentclass{PoS}
\usepackage{amsmath}

\newcommand{\be}{\begin{equation}}
\newcommand{\ee}{\end{equation}}
\newcommand{\beq}{\begin{eqnarray}}
\newcommand{\eeq}{\end{eqnarray}}

\title{Cosmology within Noncommutative Spectral Geometry}

\ShortTitle{Cosmology within  Noncommutative Spectral Geometry}

\author{\speaker{Mairi Sakellariadou}
\\
        Department of Physics, King's College, University of
  London, Strand WC2R 2LS, London, U.K.\\
        E-mail: \email{Mairi.Sakellariadou@kcl.ac.uk}}

\abstract{
Close to the Planck energy scale, the quantum nature of space-time
reveals itself and all forces, including gravity, should be unified so
that all interactions correspond to just one underlying symmetry.  In
the absence of a full quantum gravity theory, one may follow an
effective approach and consider space-time as the product of a
four-dimensional continuum compact Riemanian manifold by a tiny
discrete finite noncommutative space.  Since all available data are of
a spectral nature, one may argue that it is more appropriate to apply
the spectral action principle in this almost commutative
space. Following this procedure one obtains an elegant geometric
explanation for the most successful particle physics model, namely the
standard model (and supersymmetric extensions) of electroweak and
strong interactions in all its details, as determined by experimental
data.  Moreover, since this gravitational theory lives by construction
at very high energy scales, it offers a perfect framework to address
some of the early universe cosmological questions still awaiting for
an answer.

After introducing some of the main mathematical elements of
noncommutative spectral geometry, I will discuss various cosmological
and phenomenological consequences of this theory, focusing in
particular on constraints imposed on the gravitational sector of the
theory.
}

\FullConference{Corfu Summer Institute on Elementary Particles and
  Physics - Workshop on Non Commutative Field Theory and
  Gravity,\\ September 8-12, 2010\\ Corfu Greece}

\begin{document}

\section{Introduction}
At energies much below the Planck scale, gravity can be considered as
a classical theory, however as energies approach the Planck scale, the
quantum nature of space-time becomes apparent, and the simple
prescription, dictating that physics can be described by the sum of
the Einstein-Hilbert and the Standard Model (SM) action ceases to be
valid.  At such high energy scales, all forces, including gravity, are
expected to be unified so that all interactions correspond to one
underlying symmetry.  Thus, near Planckian energies, the appropriate
formulation of geometry should be within a quantum framework, while
the nature of space-time would change in such a way so that one can
recover the low energy picture of diffeomorphism and internal gauge
symmetries, which govern General Relativity (GR) and gauge groups on
which the Standard Model is based, respectively. A promising attempt
to obtain a quantum nature of space-time has been realised within the
realm of NonCommutative Geometry (NCG).

Noncommutative Geometry~\cite{ncg-book1, ncg-book2} is a beautiful and
rich mathematical theory, according which geometry can be described
through the functions defined on the geometry, while the geometric
properties of spaces can be described by the properties of functions
defined on the spaces. An important new feature of noncommutative
geometry is the existence of inner fluctuations of the metric, which
correspond to the subgroup of inner automorphisms. Besides the
mathematical beauty of NCG, which by itself explains why one may want
to study this theory, NCG offers a variety of phenomenological
consequences, which turn this theory into a fertile framework to
address fundamental issues of early universe cosmology and high energy
physics phenomenology.

In what follows we will follow Connes' approach~\cite{ncg-book1,
  ncg-book2} and consider a model of a two-sheeted space made from
the product of a continuous space by a discrete space. This model led
to a geometric explanation of the Standard Model; in particular the
model shows that the vacuum Expectation Value of the Higgs field is
related to the noncommutative distance between the two sheets. Within
Connes' model of NCG, the Higgs field is conformally coupled to the
Ricci curvature, while the generalised Einstein-Hilbert action
contains in addition a minimally coupled massless scalar field related
to the distance between the two sheets.  Connes' approach is based upon
a spectral action principle, stating that the bare bosonic Euclidean
action for any noncommutative model based on a product (noncommutative)
space is the trace of the heat kernel associated with the square of
the noncommutative Dirac operator of the product geometry.

Within noncommutative spectral geometry, we look for a hidden
structure in the functional of gravity coupled to the SM at today's
low energy scales, and avoid an extrapolation by many orders of
magnitude to guess the appropriate structure of space-time at
Planckian energy scales.

Noncommutative spectral geometry offers an elegant approach to
unification, based on the symplectic unitary group in Hilbert space,
rather than on finite dimensional Lie groups. The model offers a
unification of internal symmetries with the gravitational ones. All
symmetries arise as automorphisms of the noncommutative algebra of
coordinates on a product geometry.  Due to the lack of a full quantum
gravity theory, which {\sl a priori} should define the geometry of
space-time at Planckian energy scales, we will follow an effective
theory approach and consider the simplest case beyond commutative
spaces. Thus, below but close to the Planck energy scale, space-time
will be considered as the product of a Riemanian spin manifold by a
finite noncommutative space. At higher energy scales, space-time
should become noncommutative in a nontrivial way, while at energies
above the Planck scale the whole concept of geometry may altogether
become meaningless. As a next, but highly nontrivial, step one should
consider noncommutative spaces whose limit is the almost commutative
space considered here. Unfortunately, the birth of geometry may remain
an unsolved puzzle for still quite sometime.

Let me draw the attention of the reader to the fact that the
noncommutative spectral geometry approach discussed here, goes beyond
the noncommutative geometry notion employed in the literature to
implement the fuzziness of space-time by means of $[{\bf x}^i,
  {\bf x}^j]=i\theta^{ij}$, where $\theta^{ij}$ is an anti-symmetric,
real, $d\times d$ ($d$ is the dimension of space-time) matrix, and
${\bf x}^i$ denote spatial coordinates. 
\section{Elements of Noncommutative Spectral Geometry}
To extend the Riemanian paradigm of geometry to the notion of metric
on a noncommutative space, the latter should contain the Riemanian
manifold with the metric tensor (as a special case), allow for
departures from commutativity of coordinates as well as for quantum
corrections of geometry, contain spaces of complex dimension, and
offer the means of expressing the Standard Model coupled to Einstein
gravity as pure gravity on a suitable geometry.  Noncommutative
spectral geometry considers the SM as a phenomenological model, which
should however determine the geometry of space-time, so that the
Maxwell-Dirac action functional leads to the SM action. The
geometrical space turns out to be the product of a continuum manifold
for the space-time geometry and a noncommutative space for the
internal geometry of the SM. 

Adopting the simplest generalisation beyond commutative spaces, we
consider the geometry of space-time to be described by the tensor
product ${\cal M}\times{\cal F}$ of a continuum compact Riemanian
manifold ${\cal M}$ and a tiny discrete finite noncommutative space
${\cal F}$ composed of just two points.  It is worth noting that while
the metric dimension of ${\cal F}$ is zero, its $K$-theoretic
dimension is equal to 6 modulo 8.  Following this prescription, the
Lagrangian of the SM, including mixing and Majorana mass terms for
neutrinos, minimally coupled to gravity is recovered from spectral
invariants of the inner fluctuations of the product metric on ${\cal
  M}\times{\cal F}$.

The noncommutative nature of the discrete space ${\cal F}$ is given by
a spectral triple $({\cal A, H, D})$, where ${\cal A}$ is an
involution of operators on the finite-dimensional Hilbert space ${\cal
  H}$ of Euclidean fermions, and ${\cal D}$ is a self-adjoint
unbounded operator in ${\cal H}$. For commutative geometries, the
classical notion of a real variable is described as a real-valued
function on a space, described by the corresponding algebra ${\cal A}$
of coordinates, which for noncommutative geometries is represented as
operators in a fixed Hilbert space ${\cal H}$.  Since real coordinates
are represented by self-adjoint operators, all information about a
space is encoded in the algebra of coordinates ${\cal A}$, which is
related to the gauge group of local gauge transformations.  By
dropping the commutativity property, the infinitesimal line element
$ds$, employed to define geometry through the measurement of distances
$d(x,y)$ through $d(x,y)={\rm inf}\int_\gamma ds$ where the infimum is
considered over all possible paths from point $x$ to $y$, does not
need to be localised. The absence of commutation of the line element
with the coordinates renders possible the measurement of distances
through the formula
$d(x,y)={\rm sup}\{|f(x)-f(y)|:f\in {\cal A},||{\cal D},f||\leq 1\}$,
where ${\cal D}$ denotes the inverse of the line element.

Assuming the algebra ${\cal A}$ to be symplectic-unitary, implies
$\mathcal{A}=M_{a}(\mathbb{H})\oplus M_{k}(\mathbb{C})$,
with $k=2a$ and $\mathbb{H}$ denoting the algebra of quaternions,
which plays an important r\^ole here and its choice remains to be
explained; we assume quaternion linearity to obtain the SM. The choice
$k=4$ is the first value that produces the correct number ($k^2=16$)
of fermions in each of the three
generations~\cite{Chamseddine:2007ia}, with the number of generations
being a physical input. If at the Large Hadron Collider new particles
are discovered, one may be able to accommodate them by considering a
higher value for $k$.  While the choice of algebra ${\cal A}$
constitutes the main input of the theory, the choice of Hilbert space
${\cal H}$ is irrelevant, since all separable infinite-dimensional
Hilbert spaces are isomorphic.  The operator ${\cal D}$ corresponds to
the inverse of the Euclidean propagator of fermions, and is given by
the Yukawa coupling matrix which encodes the masses of the elementary
fermions and the Kobayashi--Maskawa mixing parameters.  We thus
describe geometry by focusing on the Dirac operator ${\cal D}$,
instead of the metric tensor $g_{\mu\nu}$, used for spaces whose
coordinates commute. The fermions of the SM provide the Hilbert space
${\cal H}$ of a spectral triple for the algebra ${\cal A}$, while the
bosons of the SM are obtained through inner fluctuations of the Dirac
operator of the product geometry.

All experimental data are of a spectral nature, thus our aim is to
extract information, from our noncommutative geometry construction,
which is of a spectral nature. Luckily the appropriate tool in
noncommutative geometry has been developed; the Spectral Action
Functional in noncommutative spaces is the analogous to the Fourier
Transform used in spaces characterised by commutation of the
coordinates. To obtain the full Lagrangian of the SM, minimally
coupled to gravity, we will apply the Spectral Action Principle,
stating that the bosonic part of the spectral action functional
depends only on the spectrum of the Dirac operator and its asymptotic
expression, and for large energy $\Lambda$ is of the form ${\rm
  Tr}(f({\cal D}/\Lambda))$, with $f$ being a cut-off function, whose
choice plays only a small r\^ole; both ${\cal D}$ and $\Lambda$ have
physical dimensions of a mass and there is no absolute scale on which
they can be measured. The r\^ole of the cut-off scale $\Lambda$ is
equivalent to keeping only frequencies smaller than the mass scale
$\Lambda$. According to the spectral action principle, ${\rm
  Tr}(f({\cal D}/\Lambda))$ is the fundamental action functional that
can be used not only at the classical level but also at the quantum
level, after Wick rotation to Euclidean signature.  The
cut-off-dependent Euclidean action is viewed (in the Wilsonian
approach) as the bare action at mass scale $\Lambda$.  The
physical Lagrangian has also a fermionic part, which has the simple
linear form $(1/2)\langle J\psi, {\cal D}\psi\rangle$, where $J$ is
the real structure on the spectral triple and $\psi$ are spinors
defined on the Hilbert space.  To study cosmological implications of
this gravitational model one may consider only the bosonic part of the
action; the fermionic part is important to deduce particle physics
phenomenology.

The formalism of spectral triples favours Euclidean rather than
Lorentzian signature.  The discussion of phenomenological aspects of
the theory relies on a Wick rotation to imaginary time, into the
Lorentzian signature. While sensible from the phenomenological point
of view, there exists as yet no justification on the level of the
underlying theory. It is worth noting however that the issue of
Euclidean versus Lorentzian signature is not a kind of pathology only
for the case of noncommutative spectral geometry, it is for
instance encountered in the nonperturbative path-integral approach to
quantum gravity.

Applying the spectral action principle to the inner fluctuations of
the product ${\cal M}\times{\cal F}$ of an ordinary compact spin
4-manifold with the finite noncommutative geometry, one recovers the
Standard Model action coupled to Einstein and Weyl gravity plus higher
order nonrenormalisable interactions suppressed by powers of the
inverse of the mass scale of the theory~\cite{ccm}. This model
provides specific values of some of the SM parameters at unification
scale (denoted by $\Lambda$). Following the Wilsonian approach, one
can then obtain physical predictions for the SM parameters by running
them down to low (present) energy scales through the Renormalisation
Group Equations (RGE). 

Using heat kernel methods, the trace ${\rm Tr}(f({\cal D}/\Lambda))$
can be written in terms of the geometrical Seeley - de Witt coefficients
$a_n$, which are known for any second order elliptic differential
operator, as $\sum_{n=0}^\infty F_{4-n}\Lambda^{4-n}a_n$~, where the
function $F$ is defined such that $F({\cal D}^2)=f({\cal D})$.  Thus,
the bosonic part of the spectral action can be expanded in powers of
$\Lambda$ in the form~\cite{ac1996,ac1997}
\begin{equation}
\label{eq:sp-act}
{\rm Tr}\left(f\left(\frac{{\cal D}}{\Lambda}\right)\right)\sim
\sum_{k\in {\rm DimSp}} f_{k} \Lambda^k{\int\!\!\!\!\!\!-} |{\cal
  D}|^{-k} + f(0) \zeta_{{\cal D}(0)}+ {\cal O}(1)~.
\end{equation}
The momenta $f_k$ are defined as $f_k\equiv\int_0^\infty f(u)
u^{k-1}{\rm d}u$ for $k>0$ and $f_0\equiv f(0)$, the noncommutative
integration is defined in terms of residues of zeta functions
$\zeta_{\cal D} (s) = {\rm Tr}(|{\cal D}|^{-s})$ at poles of the zeta
function, and the sum is over points in the dimension spectrum
of the spectral triple. 

Considering the Riemanian geometry to be four-dimensional, the
asymptotic expansion of the trace reads
\be\label{asymp-exp}
{\rm Tr}\left(f\left(\frac{D}{\Lambda}\right)\right)\sim
2\Lambda^4f_4a_0+2\Lambda^2f_2a_2+f_0a_4+\cdots
+\Lambda^{-2k}f_{-2k}a_{4+2k}+\cdots~.
\ee
The smooth even function $f$, which decays fast at infinity, only
enters in the multiplicative factors:
\beq
f_4=\int_0^\infty f(u)u^3 du~~~&,&~~
f_2=\int_0^\infty f(u)u du~,\nonumber\\
f_0=f(0)~~~&,&~~~
f_{-2k}=(-1)^k\frac{k!}{(2k)!} f^{(2k)}(0)~.
\eeq
Since $f$ is taken as a cut-off function, its Taylor expansion at zero
vanishes, thus its asymptotic expansion, Eq.~(\ref{asymp-exp}),
reduces to just
\be
{\rm Tr}\left(f\left(\frac{D}{\Lambda}\right)\right)\sim
2\Lambda^4f_4a_0+2\Lambda^2f_2a_2+f_0a_4~.
\ee

 The cut-off function enters through its
momenta $f_0, f_2, f_4$; these three additional real parameters are
physically related to the coupling constants at unification, the
gravitational constant and the cosmological constant.  In the
four-dimensional case, the term in $\Lambda^4$ in the spectral action,
Eq.~(\ref{eq:sp-act}), gives a cosmological term, the term in
$\Lambda^2$ gives the Einstein-Hilbert action functional with the
physical sign for the Euclidean functional integral (provided
$f_2>0$), and the $\Lambda$-independent term yields the Yang-Mills
action for the gauge fields corresponding to the internal degrees of
freedom of the metric. The scale-independent terms in the spectral
action have conformal invariance. Note that the arbitrary mass scale
$\Lambda$ can be made dynamical by introducing a scaling dilaton
field.

Since the physical Lagrangian is entirely determined by the geometric
input, the physical implications of this approach are closely
dependent on the underlying chosen geometry. The obtained physical
Lagrangian contains, in addition to the full Standard Model
Lagrangian, the Einstein-Hilbert action with a cosmological term, a
topological term related to the Euler characteristic of the space-time
manifold, a conformal Weyl term and a conformal coupling of the Higgs
field to gravity. The bosonic action in Euclidean signature
reads~\cite{ccm}
\beq\label{eq:action1} 
{\cal S}^{\rm E} = \int \left(
\frac{1}{2\kappa_0^2} R + \alpha_0
C_{\mu\nu\rho\sigma}C^{\mu\nu\rho\sigma} + \gamma_0 +\tau_0 R^\star
R^\star
\right.  
+ \frac{1}{4}G^i_{\mu\nu}G^{\mu\nu
  i}+\frac{1}{4}F^\alpha_{\mu\nu}F^{\mu\nu\alpha}\nonumber\\ 
+\frac{1}{4}B^{\mu\nu}B_{\mu\nu}
+\frac{1}{2}|D_\mu{\bf H}|^2-\mu_0^2|{\bf H}|^2
\left.
- \xi_0 R|{\bf H}|^2 +\lambda_0|{\bf H}|^4
\right) \sqrt{g} \ d^4 x~, \eeq
where 
\beq\label{bc} 
\kappa_0^2=\frac{12\pi^2}{96f_2\Lambda^2-f_0\mathfrak{c}}
~~&,&
~~\alpha_0=-\frac{3f_0}{10\pi^2}~~~,\nonumber\\ 
\gamma_0=\frac{1}{\pi^2}\left(48f_4\Lambda^4-f_2\Lambda^2\mathfrak{c}
+\frac{f_0}{4}\mathfrak{d}\right)~~&,&
~~\tau_0=\frac{11f_0}{60\pi^2}~~~,\nonumber\\ 
\mu_0^2=2\Lambda^2\frac{f_2}{f_0}-{\frac{\mathfrak{e}}{\mathfrak{a}}}~~~,
~~~\xi_0=\frac{1}{12}~~~&,&
~~~\lambda_0=\frac{\pi^2\mathfrak{b}}{2f_0\mathfrak{a}^2}~;
\eeq
${\bf H}$ is a rescaling ${\bf H}=(\sqrt{af_0}/\pi)\phi$ of the Higgs
field $\phi$ to normalize the kinetic energy, and the momentum $f_0$
is physically related to the coupling constants at unification.
Notice the absence of quadratic terms in the curvature; there is only
the term quadratic in the Weyl curvature and the topological term
$R^\star R^\star$. In a cosmological setting, namely for
Friedmann-Lema\^{i}tre-Robertson-Walker (FLRW) geometries, the Weyl
term vanishes. Notice also the term that couples gravity with the
SM, a term which should always be present when one considers gravity
coupled to scalar fields.

Writing the asymptotic expansion of the spectral action, a number of
geometric parameters appeared, which describe the possible choices of
Dirac operators on the finite noncommutative space. These parameters
correspond to the Yukawa parameters of the particle physics model and
the Majorana terms for the right-handed neutrinos. They are given
by~\cite{ccm}
\beq\label{eq:Ys-oth}
 \mathfrak{a}&=&{\rm Tr}
\left( Y^\star_{\left(\uparrow 1\right)} Y_{\left(\uparrow 1\right)} +
Y^\star_{\left(\downarrow 1\right)} Y_{\left(\downarrow
  1\right)}
+ 
3\left( Y^\star_{\left(\uparrow 3\right)} Y_{\left(\uparrow 3\right)}
+ Y^\star_{\left(\downarrow 3\right)} Y_{\left(\downarrow 3\right)}
\right)\right)~,\nonumber\\ \mathfrak{b}&=&{\rm Tr}\left(\left(
Y^\star_{\left(\uparrow 1\right)} Y_{\left(\uparrow 1\right)}\right)^2
+ \left(Y^\star_{\left(\downarrow 1\right)} Y_{\left(\downarrow
  1\right)}\right)^2
+ 
3\left( Y^\star_{\left(\uparrow 3\right)} Y_{\left(\uparrow
  3\right)}\right)^2 + 3 \left(Y^\star_{\left(\downarrow 3\right)}
Y_{\left(\downarrow 3\right)}
\right)^2\right)~,\nonumber\\ \mathfrak{c}&=&{\rm Tr}\left(Y^\star_R
Y_R\right)~,\nonumber\\ \mathfrak{d}&=&{\rm Tr}\left(\left(Y^\star_R
Y_R\right)^2\right),\nonumber\\ \mathfrak{e}&=&{\rm Tr}\left(Y^\star_R
Y_RY^\star_{\left(\uparrow 1\right)} Y_{\left(\uparrow
  1\right)}\right)~, \eeq
with $Y_{\left(\downarrow 1\right)}, Y_{\left(\uparrow 1\right)},
Y_{\left(\downarrow 3\right)}, Y_{\left(\uparrow 3\right)}$ and $Y_R$
being $(3\times 3)$ matrices, with $Y_R$ symmetric. The $Y$ matrices
are used to classify the action of the Dirac operator and give the
fermion and lepton masses, as well as lepton mixing, in the asymptotic
version of the spectral action.  The Yukawa parameters run with the
RGE of the particle physics model. Since running towards lower
energies implies that nonperturbative effects in the spectral action
cannot be any longer neglected, any results based on the asymptotic
expansion and on renormalisation group analysis can only hold for
early universe cosmology.  Hence, the spectral action at the
unification scale $\Lambda$ offers a framework to investigate early
universe cosmological
models~\cite{Nelson:2008uy,Nelson:2009wr,Marcolli:2009in,mmm,Nelson:2010ru,Nelson:2010rt}. For
later times, one should consider the full spectral action, a direction
which requires the development of nontrivial mathematical tools.

It is important to emphasise that the relations given in
Eq.~(\ref{bc}) above are tied to the scale at which the expansion is
performed.  There is {\sl a priori} no reason for the constraints to
hold at scales below the unification scale $\Lambda$, since they
represent mere boundary conditions.  One should therefore be very
careful and keep in mind that it is incorrect to consider the
relations in Eq.~(\ref{bc}) as functions of the energy scale; these
relations are only valid at unification scale $\Lambda$.

\section{High Energy Phenomenology of the Noncommutative Spectral Geometry}  
In what follows we assume that the function $f$ is well approximated
by the cut-off function and ignore higher order terms.  Normalisation
of the kinetic terms implies
\be
\frac{g_3^2f_0}{ 2\pi^2}=\frac{1}{4} ~~\mbox{and}~~ g_3^2=g_2^2=
\frac{5}{ 3}g_1^2\nonumber~, \ee
while
\be \sin^2\theta_{\rm W}=\frac{3}{8}~; \ee
a relation which was also found for SU(5) and SO(10).  Assuming the
big desert hypothesis, the running of the couplings
$\alpha_i=g_i^2/(4\pi)$ with $i=1,2,3$ can then be obtained via the
RGE.

The phenomenological consequences of the noncommutative spectral
geometry as an approach to unification have been discussed in
Ref.~\cite{ccm}, where the authors considered only one-loop corrections,
for which the $\beta$-functions are $\beta_{g_i}=(4\pi)^{-2}b_ig_i^3$
with $i=1,2,3$ and $b=(41/6,-19/6,-7)$. It is worth noticing that only
at one-loop order the Renormalisation Group Equations for the coupling
constants $g_i$ are uncoupled from the other Standard Model parameters.

Performing one-loop RGE for the running of the gauge couplings and the
Newton constant, it was shown~\cite{ccm} that these do not meet at a
point, the error being within just few percent. The fact that the
predicted unification of the coupling constants does not hold exactly,
implies that the big desert hypothesis is only approximately
valid and new physics are expected between unification and
present energy scales. In terms of our assumption for the cut-off
function, the lack of a unique unification energy, implies that even
though the function $f$ can be approximated by the cut-off function,
there exist small deviations.  Besides this result it is however worth
noting that the model leads to the correct representations of the
fermions with respect to the gauge group of the SM, the Higgs doublet
appears as part of the inner fluctuations of the metric, and
Spontaneous Symmetry Breaking mechanism arises naturally with the
negative mass term without any tuning. In addition, the see-saw
mechanism is obtained, the 16 fundamental fermions are recovered, and
a top quark mass of $M_{\rm top}\sim 179 ~{\rm GeV}$ is predicted.

The model predicts a heavy Higgs mass; in zeroth order approximation,
it predicts a mass of the Higgs boson approximately equal to $170 ~{\rm
  GeV}$, which strictly speaking is ruled out by current experimental
data.  Due to this discrepancy between the NCG prediction and the
experimental data, noncommutative spectral geometry has been (rather
unfairly) criticised, even though the result quoted above depends on
the value of the gauge couplings at unification scale, which is
uncertain and was obtained neglecting the nonminimal coupling between
the Higgs field and the Ricci curvature. I believe that one should
instead draw the conclusion that noncommutative spectral geometry as
an approach to unification, even in its present (and certainly
simple) version, it still led to the correct order of magnitude for
the Higgs mass, a result which was by no means obvious.

Considering an energy scale $\Lambda\sim 1.1\times 10^{17}\ {\rm
  GeV}$, the standard form of the gravitational action and the
experimental value of Newton's constant at ordinary scales imply
 $\kappa_0^{-1}\sim 2.43\times 10^{18}\ {\rm GeV}$.

Finally, this approach to unification does not provide any explanation
of the number of generations, nor leads to constraints on the values
of the Yukawa couplings.

\section{Cosmological Consequences of the Noncommutative Spectral Geometry}
The Lorentzian version of the gravitational part of the asymptotic
formula for the bosonic sector of the action obtained within
noncommutative spectral geometry reads~\cite{ccm}
\be\label{eq:1.5} {\cal S}_{\rm grav}^{\rm L} = \int \left(
\frac{1}{2\kappa_0^2} R + \alpha_0
C_{\mu\nu\rho\sigma}C^{\mu\nu\rho\sigma} + \tau_0 R^\star
R^\star
\xi_0 R|{\bf H}|^2 \right)
\sqrt{-g} \ d^4 x~. \ee
This action leads to the following equations of
motion~\cite{Nelson:2008uy}
\be\label{eq:EoM2} R^{\mu\nu} - \frac{1}{2}g^{\mu\nu} R +
\frac{1}{B^2} \delta_{\rm cc}\left[
  2C^{\mu\lambda\nu\kappa}_{;\lambda ; \kappa} +
  C^{\mu\lambda\nu\kappa}R_{\lambda \kappa}\right]
\nonumber\\ = 
\kappa_0^2 \delta_{\rm cc}T^{\mu\nu}_{\rm matter}~, \ee
where we have introduced $B^2 \equiv -(4\kappa_0^2
\alpha_0)^{-1}$, related to the $f_0$ moment of the cut-off function
and we have captured the nonminimal coupling between the Higgs field
and the Ricci curvature scalar in the parameter $\delta_{\rm cc}$,
defined as $\delta_{\rm cc}\equiv[1-2\kappa_0^2\xi_0{\bf H}^2]^{-1}.$

In the low energy weak curvature regime, the nonminimal coupling
between the background geometry and the Higgs field can be neglected,
thus $\delta_{\rm cc}=1$.  For a FLRW space-time, the Weyl tensor
vanishes, hence the NCG corrections to the Einstein equation
vanish~\cite{Nelson:2008uy}, rending difficult to restrict $B$ via
cosmology or solar-system tests. Imposing a lower limit on $B$
would imply an upper limit to the moment $f_0$, and thus restrict
particle physics at unification.

One can impose an upper limit to the moment $f_0$, by studying the
energy lost to gravitational radiation by orbiting
binaries~\cite{Nelson:2010ru,Nelson:2010rt}.  Considering linear
perturbations around a Minkowski background metric, the equations of
motion read~\cite{Nelson:2010rt}
\be\label{eq:1} \left( \Box - B^2 \right) \Box h^{\mu\nu} =
B^2 \frac{16\pi G}{c^4} T^{\mu\nu}_{\rm matter}~, \ee 
where $T^{\mu\nu}_{\rm matter}$ is taken to lowest order in
$h^{\mu\nu}$. Since $B$ plays the r\^ole of a mass, it must be
real and positive, thus $\alpha_0$ must be negative for Minkowski
space to be a stable vacuum of the theory.

Consider the energy lost to gravitational radiation by orbiting
binaries. In the far field limit, $|{\bf r}| \approx |{\bf r} - {\bf
  r}'|$ (${\bf r}$ and ${\bf r}'$ denote the locations of observer
and emitter, respectively), the spatial components of the general
first order solution for a perturbation against a Minkowski background
read~\cite{Nelson:2010rt}
\be\label{eq:4} h^{ik}\left( {\bf r},t\right) \approx \frac{2G
  B}{3c^4} \int_{-\infty}^{t-\frac{1}{c}|{\bf r}|} \frac{d
  t'}{\sqrt{c^2\left( t-t'\right)^2 - |{\bf r}|^2} }
          {\cal J}_1 \left( B\sqrt{c^2\left( t-t'\right)^2 - |{\bf
              r}|^2}\right) \ddot{D}^{ik}\left(t'\right)~; \ee
$D^{ik}$ is the quadrupole moment, defined as $D^{ik}\left(t\right)
\equiv \frac{3}{c^2}\int x^i x^k T^{00}({\bf r},t) \ d{\bf r}$, and
${\cal J}_1$ a Bessel function of the first kind. While for
$B\rightarrow \infty$ the theory reduces to GR,
for finite $B$ gravitational radiation contains massive and
massless modes, both sourced from the quadrupole moment of the system.

For a binary pair of masses $m_1, m_2$ in circular orbit in the
$(xy)$-plane, the rate of energy loss is
\be\label{eq:energy} -\frac{{\rm d} {\cal E}}{{\rm d}t} \approx
\frac{c^2}{20G} |{\bf r}|^2 \dot{h}_{ij} \dot{h}^{ij}~,  \ee
with~\cite{Nelson:2010rt}
\be
\dot{h}^{ij}\dot{h}_{ij}= \frac{128\mu^2|\rho|^4 \omega^6 G^2
  B^2}{c^8}
 \times \left[ f_{\rm c}^2\left(B|{\bf
    r}|,\frac{2\omega}{B c}\right) + f_{\rm s}^2\left(B|{\bf
    r}|,\frac{2\omega}{B c}\right)\right]~, \ee
\beq\label{eq:f1}
 f_{\rm s}\left( x,z\right) &\equiv& \int_0^\infty
\frac{d s}{\sqrt{s^2 + x^2}} {\cal J}_1\left(s\right) \sin
\left(z\sqrt{ s^2 + x^2} \right)~,\\
\label{eq:f2}
f_{\rm c}\left( x,z\right) &\equiv&
\int_0^\infty \frac{d s}{\sqrt{s^2 + x^2}} {\cal
  J}_1\left(s\right) \cos \left(z\sqrt{ s^2 + x^2} \right)~.
\eeq
The orbital frequency $\omega$, defined in terms of the magnitude
$|\rho|$ of the separation vector between the two bodies, is constant
and equal to $\omega = |\rho|^{-3/2} \sqrt{ G\left( m_1 +
  m_2\right)}$.

The integrals in Eqs.~(\ref{eq:f1}), (\ref{eq:f2}), exhibit a strong
resonance behavior at $z=1$, which corresponds to the critical
frequency~\cite{Nelson:2010rt}
\be
\label{critical}
2\omega_{\rm c} =B c~,
\ee
around which strong deviations from the GR results are expected.  This
maximum frequency results from the natural length scale, given by
$B^{-1}$, at which NCG effects become dominant.

There are several binary pulsars for which the rate of change of the
orbital frequency has been well characterised, and the predictions of
General Relativity agree with the data to a high accuracy. Requiring the
magnitude of deviations from GR obtained in the
context of noncommutative spectral geometry, to be less than the
allowed uncertainty in the data,  one
constrains $B$, namely~\cite{Nelson:2010ru}
\be 
\label{constr-beta}
B > 7.55\times 10^{-13}~{\rm m}^{-1}~.
\ee

This observational constraint may seem weak, however it is comparable
to existing constraints on similar, {\sl ad hoc}, additions to GR. In
particular, constraints on additions to the Einstein-Hilbert action,
of the form $R^2$ and $R_{\mu\nu} R^{\mu\nu}$, are of the order of
$B_{R^2} \geq 3.2\times 10^{-9} {\rm m}^{-1}$, where $B_{R^2}$ is the
$B$ parameter associated with the couplings of these
terms~\cite{Stelle}.  Note also that since the strongest constraint
comes from systems with high orbital frequencies, future observations
of rapidly orbiting binaries, relatively close to the Earth, could
improve it by many orders of magnitude.

Remaining in the low-energy limit, in other words neglecting the
coupling between the Higgs field and the background geometry, on may
consider the corrections to the background Einstein's equations. It
turns out that noncommutative corrections do not occur at the level of
a FLRW background, since then the modified Friedmann equation reduces
to its standard form~\cite{Nelson:2008uy}. One may have naively
claimed that this was expected, arguing that in a spatially
homogeneous space-time the spatial points are equivalent and any
noncommutative effects are then expected to vanish. However, this
argument does not apply here, since the noncommutativity is
incorporated in the internal manifold ${\cal F}$ and the space-time is
a commutative four-dimensional manifold.

Neglecting the nonminimal coupling between the Higgs field and
the Ricci curvature, any modifications to the background equation will
be apparent at leading order for anisotropic and inhomogeneous models.
Let us consider the representative example of Bianchi type-V model,
for which the space-time metric, in Cartesian coordinates, reads
\be g_{\mu\nu} = {\rm diag} \left[ -1,\{a_1(t)\}^2e^{-2nz} ,
 \{a_2(t)\}^2e^{-2nz}, \{a_3(t)\}^2 \right]~; \ee
$a_i(t)$ with $i=1,2,3$ arbitrary functions and $n$ is an integer.

Defining
$A_i\left(t\right) = {\rm ln} a_i\left(t\right)$  with $i=1,2,3$,
the modified Friedmann equation reads~\cite{Nelson:2008uy}:
\beq\label{eq:Friedmann_BV} \kappa_0^2 T_{00}=&&\nonumber\\
 - \dot{A}_3\left(
\dot{A}_1+\dot{A}_2\right) -n^2 e^{-2A_3} \left( \dot{A}_1
\dot{A}_2-3\right)&& \nonumber \\
 +\frac{8\alpha_0\kappa_0^2 n^2}{3} e^{-2A_3} \left[
  5\left(\dot{A}_1\right)^2 + 5\left(\dot{A}_2\right)^2 -
  \left(\dot{A}_3\right)^2\right.
\left. 
- \dot{A}_1\dot{A}_2 - \dot{A}_2\dot{A}_3
  -\dot{A}_3\dot{A}_1 - \ddot{A}_1 - \ddot{A}_2 - \ddot{A}_3 + 3
  \right]
&& \nonumber \\
- \frac{4\alpha_0\kappa_0^2}{3} \sum_i \Biggl\{
\dot{A}_1\dot{A}_2\dot{A}_3 \dot{A}_i
 + \dot{A}_i \dot{A}_{i+1} \left(
\left( \dot{A}_i - \dot{A}_{i+1}\right)^2 -
\dot{A}_i\dot{A}_{i+1}\right)
&& \nonumber \\
 + \left( \ddot{A}_i + \left( \dot{A}_i\right)^2\right)\left[
  -\ddot{A}_i - \left( \dot{A}_i\right)^2 + \frac{1}{2}\left(
  \ddot{A}_{i+1} + \ddot{A}_{i+2} \right)
\right.
\left. 
+ \frac{1}{2}\left(
  \left(\dot{A}_{i+1}\right)^2 + \left( \dot{A}_{i+2}\right)^2 \right)
  \right]
&& \nonumber \\ 
+ \left[ \dddot{A}_i + 3 \dot{A}_i \ddot{A}_i -\left(\ddot{A}_i +
  \left(\dot{A}_i\right)^2 \right)\left( \dot{A}_i - \dot{A}_{i+1} -
  \dot{A}_{i+2} \right)\right]
\left[ 2\dot{A}_i
  -\dot{A}_{i+1}-\dot{A}_{i+2} \right]\Biggr\} \eeq
with $i=1,2,3$, while the $t$-dependence of the terms has been omitted
for simplicity.  

Any term containing $\alpha_0$ in Eq.~(\ref{eq:Friedmann_BV}), encodes
a modification from the standard result.  Let us study
Eq.~(\ref{eq:Friedmann_BV}) above. The correction terms can be divided
into two types.  The first one contains the terms in braces in
Eq.~(\ref{eq:Friedmann_BV}), which are fourth order in time
derivatives. Thus, for the slowly varying functions, usually
considered in cosmology, these corrections can be neglected.  The
second type, which appears in the third line in
Eq.~(\ref{eq:Friedmann_BV}), occurs at the same order as the standard
Einstein-Hilbert terms. However, since this term is proportional to
$n^2$, it vanishes for homogeneous versions of Bianchi type-V.  Thus,
although anisotropic cosmologies do contain corrections due to the
additional NCG terms in the action, they are typically of higher
order~\cite{Nelson:2008uy}.  Inhomogeneous models do contain
correction terms that appear on the same footing as the original
terms.

It is worth noting that by studying the case of the Bianchi type~V
model, one can identify the noncommutative geometry effects in other
cases of cosmological models, for instance Bianchi~I and Kasner
models.  In conclusion, the corrections to Einstein's equations can
only be important for inhomogeneous and anisotropic
space-times~\cite{Nelson:2008uy}.

Certainly, the coupling between the Higgs field and the background
geometry cannot be neglected once the energies reach the Higgs
scale. The nonminimal coupling of Higgs field
to curvature leads to corrections to Einstein's equations even for
homogeneous and isotropic cosmological models. To illustrate the
effects of these corrections let us neglect the conformal term in
Eq.~(\ref{eq:EoM2}), so that the equations of motion
read~\cite{Nelson:2008uy}
\be R^{\mu\nu} - \frac{1}{2}g^{\mu\nu}R =
\kappa_0^2\left[\frac{1}{1-\kappa_0^2 |{\bf H}|^2/6}\right]
T^{\mu\nu}_{\rm matter}~, \ee
implying that $|{\bf H}|$ plays the r\^ole of an effective
gravitational constant~\cite{Nelson:2008uy}.

The nonminimal coupling between the Higgs field and the Ricci
curvature may turn out to be crucial in early universe
cosmology~\cite{Nelson:2009wr,mmm}.  Such a coupling has been
introduced {\sl ad hoc} in the literature, in an attempt to drive
inflation through the Higgs field, and thus cure one of the main
pathologies of the inflationary paradigm, namely the origin of the
inflaton field.  However, the value of the coupling constant between
the scalar field and the background geometry should be dictated by the
underlying theory.  Actually, even if classically the coupling between
the Higgs field and the Ricci curvature could be set equal to zero, a
nonminimal coupling will be induced once quantum corrections in the
classical field theory are considered.

In a FLRW metric, the Gravity-Higgs sector of the asymptotic
expansion of the spectral action, in Lorentzian
signature, reads
\be
S^{\rm
  L}_{\rm GH}=\int\Big[\frac{1-2\kappa_0^2\xi_0
    H^2}{2\kappa_0^2}R 
-\frac{1}{2}(\nabla  H)^2- V(H)\Big] \sqrt{-g}\  d^4x~,
\ee
where 
\be\label{higgs-pot}
V(H)=\lambda_0H^4-\mu_0^2H^2~,
\ee
with $\mu_0$ and $\lambda_0$ subject to radiative corrections as
functions of energy.  For large enough values of the Higgs field, the
renormalised value of $\mu_0$ and $\lambda_0$ must be calculated. More
precisely, one takes the measured values of the gauge couplings at low
energy and using $\beta$-functions one evolves them in higher energy
scales, taking into account the thresholds where quark species come
into the running. Note that one should evolve simultaneously the
running of the top Yukawa coupling and the gauge couplings.

At high energies the mass term in Eq.~(\ref{higgs-pot}) is
sub-dominant and can be neglected. As an explicit analysis has
shown~\cite{mmm}, for each value of the top quark mass there is a
value of the Higgs mass where the effective potential is about to
develop a metastable minimum at large values of the Higgs field and
the Higgs potential is locally flattened.  Calculating~\cite{mmm} the
renormalisation of the Higgs self-coupling up to two-loops, we have
constructed an effective potential which fits the renormalisation
group improved potential around the flat region.  We have
found~\cite{mmm} a very good analytic fit to the Higgs potential
around the minimum of the potential:
\be
V^\text{eff}=\lambda_0^\text{eff}(H)H^4
=[a\ln^2(b\kappa H)+c] H^4~,
\ee
where the parameters $a, b$ are related to the low energy values of
top quark mass $m_{\rm t}$ as~\cite{mmm}
\beq
a(m_\text{t})&=&4.04704\times10^{-3}-4.41909\times10^{-5}
\left(\frac{m_\text{t}}{\text{GeV}}\right)
+1.24732\times10^{-7}\left(\frac{m_\text{t}}{\text{GeV}}\right)^2~,
\nonumber\\ 
b(m_\text{t})&=&\exp{\left[-0.979261
\left(\frac{m_\text{t}}{\text{GeV}}-172.051\right)\right]}~.
\eeq
The third parameter, $c$, encodes the appearance
of an extremum and depends on the values for top quark mass and Higgs
mass.  An extremum occurs if and only if $c/a\leq 1/16$, the
saturation of the bound corresponding to a perfectly flat region.  
It is convenient to write $c=[(1+\delta)/16]a$, where $\delta=0$
saturates the bound below which a local minimum is formed.  

This study was done in the case of minimal coupling, while the
modifications for $\xi_0=1/12$ imply that flatness does not occur at
$\delta=0$, but for fixed values of $\delta$ depending on the value of
the top quark mass. Hence, for inflation to occur via the Higgs field,
the top quark mass fixes the Higgs mass extremely accurately.

Since the region where the potential is flat is narrow, to achieve a
long enough period of quasi-exponential expansion, requires that
slow-roll must be indeed very slow.  Thus, the slow-roll parameters,
$\epsilon$ and $\eta$ must be slow enough to allow sufficient number
of e-folds. In addition, the amplitude of density perturbations
$\Delta_\mathcal{R}^2$ in the Cosmic Microwave Background (CMB) must
be within the window allowed from the most recent experimental
data. More precisely, inflation predicts that at horizon crossing
(denoted by stars), the amplitude of density perturbations is related
to the inflaton potential through $\left(V_*/\epsilon_*\right)^{1/4}
=2\sqrt{3\pi}\ m_\text{Pl}\ \Delta_\mathcal{R}^{1/2}$, where
$\epsilon_*\leq1$.  Its value, as measured by
WMAP7~\cite{Larson:2010gs}, requires
$\left(V_*/\epsilon_*\right)^{1/4} =(2.75\pm0.30)\times
10^{-2}\ m_\text{Pl}$, where $m_\text{Pl}$ is the Planck mass.

A systematic search in the parameter space was performed in
Ref.~\cite{mmm} using a Monte-Carlo chain.  It was found that even
though slow-roll inflation can be realised -- a result which does not
hold for minimally coupled Higgs field -- the resulting ratio of
perturbation amplitudes is too large for any experimentally allowed
values for the masses of the top quark and the Higgs boson.  Hence,
Higgs driven inflation cannot be accommodated in the noncommutative
spectral geometry model studied here, a result which coincides with
the conclusion known for ordinary commutative geometries.  It is worth
noting that running of the gravitational constant and corrections by
considering the more appropriate de\,Sitter space-time, instead of the
Minkowski geometry employed here, do not improve substantially the
realisation of a successful slow-roll inflationary era~\cite{mmm}.

The NCG Spectral Action provides, in addition to the Higgs field,
another conformally coupled (massless) scalar field, which exhibits no
coupling to the matter sector~\cite{Chamseddine:2009yf}. Including
this field, the cosmologically relevant terms in the Wick rotated
action read
\be
S^{\rm L}=\int\left\{\frac{1}{2\kappa^2}
R - \xi_{\bf H} R{\bf H}^2 - \xi_\sigma R \sigma^2 
-\frac{1}{2}(\nabla {\bf H})^2  -\frac{1}{2}(\nabla \sigma)^2 
- V({\bf H},\sigma)\right\}\sqrt{-g}d^4x~,
\ee
where
\be V({\bf H},\sigma)=\lambda_{\bf H}{\bf H}^4-\mu_{\bf H}^2{\bf
  H}^2+\lambda_\sigma \sigma^4+\lambda_{{\bf H}\sigma}|{\bf
  H}|^2\sigma^2~.  \ee
The constants are related to the underlying parameters as
\be
\xi_{\bf H} =\frac{1}{12}~~,~~\xi_\sigma
=\frac{1}{12}~~,~~\lambda_{\bf H}
=\frac{\pi^2\mathfrak{b}}{2f_0\mathfrak{a}^2}~~,~~
\lambda_\sigma =
\frac{\pi^2\mathfrak{d}}{f_0\mathfrak{c}^2}~~,~~\mu_{\bf H}
=2\Lambda^2\frac{f_2}{f_0}~~,~~\lambda_{{\bf
    H}\sigma}=\frac{2\pi^2\mathfrak{e}}{a\mathfrak{c}f_0}~.
\ee
A careful analysis performed in Ref.~\cite{mmm} has shown that neither
this field can lead to a successful slow-roll inflationary era
if the coupling values are conformal.

It is worth noting that the above conclusions may alter if
$\xi_0=1/12$ turns out not to be a generic feature of noncommutative
geometries. However, let me emphasise that there are no nonconformal
values for the coupling $\xi_0$ for which there is a renormalization
group flow towards the conformal value as one runs the Standard Model
parameters up in the energy scale. This implies that if one
expects an exactly conformal coupling for the Higgs field at some
specific scale, it will be exactly conformal at all scales.

\section{Outlook and Conclusions}
Noncomutative spectral geometry provides an elegant way of expressing
the full Standard Model of strong and electroweak interactions coupled
to Einstein gravity, as pure gravity on a modified space-time
geometry. The paradigm of metric noncommutative geometry studied here
is of a spectral nature, an important notion in physics since all
experimental data are indeed of a spectral type. This approach is
fundamentally different than any other paradigm, which imposes a
particular structure for geometrical spaces in the quantum gravity
regime. Connes' model focuses on an almost commutative space,
considering that at energies close but lower than Planckian energy
scales, space can be described by the tensor product of a continuum
manifold by a discrete space.

Within the context of noncommutative spectral geometry, gravity and
matter are treated in a similar way, leading to concrete relationships
between matter and gravitational couplings. The asymptotic expansion
of the gravitational sector of the theory leads to modifications to
General Relativity which can be used to constrain the theory through
astrophysical observations.  Considering the energy lost by binary
systems to gravitational radiation, we were able to restrict the value
of the Weyl squared coupling in the bosonic action. 

Investigating Higgs driven inflation within noncomutative spectral
action, we have shown that while the Higgs potential can lead to the
slow-roll conditions being satisfied once the running of the
self-coupling at two-loops is included, the constraints imposed from
the CMB data make the predictions of such a scenario incompatible with
the measured value of the top quark mass. Another massless scalar
field, which naturally appears in the model, seems also not to lead to
a successful era of slow-roll inflation. However, the arbitrary mass
scale $\Lambda$ in the spectral action for the Dirac operator can be
made dynamical by introducing a dilaton field; this dilaton field may
turn out to be a successful inflaton candidate.

Noncommutative spectral geometry faces, to my opinion, at least two
immediate research directions, essential in order to deduce further
cosmological and phenomenological consequences of this
paradigm. Firstly, one should compute higher order terms in the
asymptotic expansion of the spectral action functional.  Note that it
is very difficult to compute exactly the spectral action in its
nonperturbative form, even though some progress has been made however
recently~\cite{nonpert}. Since the action functional ${\rm Tr}(f
({\cal D}/\Lambda))$ is not local -- its locality is only achieved
when it is replaced by the asymptotic expansion -- at least the next
term in the asymptotic expansion must be computed, in order to check
the validity of the asymptotic expansion.  It was recently
shown~\cite{nonpert} that for a space-time whose spatial sections are
3-spheres $S^3$, Wick rotated and compactified to a Euclidean model
$S^3 \times S^1$, the spectral action is given, for any test function,
by the sum of two terms up to a remarkably tiny correction. Let me
emphasise that for any low-energy astrophysical consequence of the
noncommutative spectral geometry, {\sl a priori} the full spectral
action, and not only its asymptotic form, has to be considered.

Secondly, it could be of great importance to find the running of the
parameters appearing in the spectral action, since otherwise it is
impossible to extract information for low-energy astrophysical
events. It is worth repeating that the expressions for $\kappa_0,
\alpha_0, \gamma_0, \tau_0, \mu_0, \lambda_0$ in terms of $f_0, f_2,
f_4, \mathfrak{a}, \mathfrak{b}, \mathfrak{c}, \mathfrak{d},
\mathfrak{e}$ and the conformal value for $\xi_0$ are only valid at
unification scale $\Lambda$. It is simply incorrect to naively
postulate that these equalities can hold at lower energy scales as
such, by just considering the parameters $\kappa, \alpha, \cdots$ as
functions of the energy scale.

At last but not least, one should consider less trivial noncommutative
spaces whose limit is the almost commutative space considered in the
original Connes' model discussed here.

Nevertheless, besides these necessary further developments, it is fair
to conclude by stating that noncommutative spectral geometry offers a
beautiful mathematical construction which provides an elegant
explanation for the most successful particle physics model at hand.

\vskip1.truecm
It is a pleasure to thank the organisers of the Workshop on Non
Commutative Field Theory and Gravity, held in the beautiful island of
Corfu, for inviting me to present this work during a stimulating and
interesting meeting.
\vskip1.truecm

\end{document}